\documentclass[aps,twocolumn,showpacs,floatfix]{revtex4}

\usepackage{graphicx,bm}
\begin{document}
\title{Singularity formation in the
 Gross-Pitaevskii Equation and Collapse in BEC}

\author{A.V. Rybin${}^{\dag\ast}$, I.P. Vadeiko${}^{\dag}$, G.G. Varzugin${}^{\ddagger}$ and J.
Timonen${}^{\dag}$} \affiliation{{\dag Department of Physics,
University of Jyv{\"a}skyl{\"a}}\\
 {PO Box 35, FIN-40351}
{Jyv{\"a}skyl{\"a}, Finland}}\email{rybin@phys.jyu.fi}
\affiliation{{$\ddagger$ Institute of Physics}
 {St. Petersburg State University}\\
 {198904, St. Petersburg, Russia}}

\begin{abstract}
We study a mechanism of  collapse of the   condensate wave
function in the Gross-Pitaevskii theory  with attractive
interparticle interaction.   We reformulate the Gross-Pitaevskii
equation as Newton's equations for the particle flux   and
introduce a collapsing fraction of particles. We assume that the
collapsing fraction is expelled from the condensate due to
dissipation. Using this hypothesis we analyze the dependence of
the condensate collapse on the initial conditions. We found that
for a properly chosen negative scattering length the remnant
fraction becomes larger  when the initial aspect ratio is
increased.
\end{abstract}
 \maketitle

\section{Introduction}
Collapse phenomena in Bose-Einstein condensates received a great
deal of interest in the last few years
Refs.\cite{pit,x,xx,xxx,xxxx,xxxxx,xxxxxx,sackett}.  The
experimental technique based on a Feshbach resonance allows to
tune,
  through application of an external magnetic field, the strength of the interatomic
interaction  and even switch between repulsive and attractive
interactions\cite{cornish}.     The realization of a collapse
controlled by the magnetic field on a gas of ${}^{85}\mbox{Rb}$
reveals a rich dynamical properties of the collapsing condensates
Ref.\cite{donley2001}.

When the number of atoms becomes sufficiently large,  the
interatomic energy overcomes the quantum pressure and the
condensate implodes. In the course  of the implosion stage the
density increases in the small vicinity of the trap center. When
the density approaches certain critical values a fraction of the
atoms gets expelled. In a time period of an order of few
milliseconds  the condensate stabilizes. There are two observable
components in the final stage of the condensate dynamics: {\it
remnant} and {\it burst} atoms. The remnant atoms are the atoms
which remain in the condensate. The burst atoms have the energy
which is much larger then the energy of the condensed atoms. There
is also a fraction of atoms, which is not observable. This
fraction is referred to as {\it missing} atoms. To account for
dissipative losses in the collapse phenomena, the most part of
theoretical studies reflected in existing literature try to modify
the mean field theory through addition of damping terms modelling
the  mechanism of dissipative losses in the system. In particular,
the results of experiment of Ref.\cite{donley2001} have been
analyzed by numerical solution of the Gross-Pitaevskii (GP)
equation with three body recombination term Refs.\cite{saito2001,
santos2001,abhi2002}. A mechanism of losses through elastic
collisions was discussed in Ref.\cite{stoof}.

The References~\cite{saito2001, santos2001} follow the ideas of
experimental work ~\cite{donley2001} and obtain the burst atoms
fraction fitting the Gaussian distribution to the condensate wave
function.  The essence of the approach of Refs.~\cite{saito2001,
santos2001} is the idea that the final state of the condensate is
comprised of the remnant fraction surrounded by a very dilute
fraction of burst atoms. Hence, the only fraction which is
expelled from the condensate is that of the missing atoms, while
the burst atoms fraction is formed in the course of the condensate
dynamics.

Effective workings of a nonlinear mechanism of dissipation in
description of collapse in BEC are inherently connected to the
structure of singularity~\cite{zakh1991} which is  formed in the
course of the nonlinear dynamics as described by  the
Gross-Pitaevskii equation~\cite{pit}. This equation reads
\begin{equation}
i \hbar\Psi_t + \frac{\hbar^2}{2m}\Delta\Psi-g\Psi|\Psi|^2-
V({{\bf x}})\Psi=0.\label{GP}
\end{equation}
Here $\Psi({\bf x},t)$ is the wave function of the condensate, the
external potential $V({\bf x})$ models the wall-less confinement
(the trap), $m$ is the mass of an individual atom,
$g=4\pi\hbar^2m^{-1}a_s$ is the effective interaction strength
(the 'coupling constant'), $a_s$ is
 the scattering length, and
$\Delta=\sum_i\frac{\partial^2}{\partial x_i^2}$ is the Laplace
operator. A convenient as well as practical choice for the
confining trap is the paraboloidal potential
$V=\frac{m}{2}\sum_i^3\omega_i^2 {x_i}^2$. Notice that so far very
little is known about the singularity structure of the $3D$ GP
equation Eq.(\ref{GP}) outside  treatments of spherically
symmetric case (see the monograph~\cite{sulem} and references
therein). In this work we will try to gain an insight into the
problem through reformulation of the Gross-Pitaevskii equation
Eq.(\ref{GP}) in a different form. Our approach resembles, at
least formally, the de Broglie-Bohm formulation of quantum
mechanics~\cite{holland,bohm}.

We introduce quantum trajectories ($q$-trajectories) of the GP
equation~Eq.(\ref{GP}). These trajectories satisfy a system of
ordinary differential equations, viz.
\begin{equation}
\frac{\partial r_i}{\partial
t}=\frac{1}{m}\frac{\partial\phi}{\partial x_i}(\mathbf{r}(
\bm{\eta},t),t);\;\;\;r_i( \bm{\eta},0)=\eta_i,\label{Eqforr}
\end{equation}
with $i=1,2,3$ in the $3D$ case. Here  $\phi$ is the phase defined
by the representation $\Psi=\sqrt{\rho} e^{i\phi/\hbar}$ of the
condensate field. The fields $r_i( \bm{\eta},t)$ can now be
interpreted as trajectories of the fictitious particles with the
initial points located at $\eta_i$. These trajectories satisfy the
Newton equations of motion for a set of particles,
\begin{equation}
m\frac{\partial^2r_i}{\partial t^2}=-\frac{\partial V}{\partial
x_i}-\frac{\partial U}{\partial x_i}, \label{EqNewton}
\end{equation}
 where the 'pressure potential' $U({\bf x},t)$ is
\begin{equation}
\label{pressures} U({\bf
x},t)=-\frac{\hbar^2}{2m}\rho^{-1/2}\Delta\rho^{1/2}+ g\rho.
\end{equation} In addition, the Liouville formula for
Eq.(\ref{Eqforr}) can be written in the form
\begin{equation}
\det\left(\frac{\partial
r_i}{\partial\eta_k}\right)\rho(\mathbf{r},t)=\rho_0( \bm{\eta}),
\label{relation}
\end{equation} where $\rho_0$ is the initial density of the
condensate. The system of equations Eqs.(\ref{EqNewton}),
(\ref{relation}) is equivalent to the original GP equation
Eq.(\ref{GP}).

As was indicated above the $q$-trajectories are used in the de~
Broglie-Bohm causal interpretation of quantum mechanics
Refs.\cite{holland,bohm}. In our previous work
Ref.\cite{rybin2001} we employed these trajectories to define
singularities of the GP equation Eq.(\ref{GP}). Indeed,
Eq.(\ref{relation})  immediately suggests that a set of singular
points $\mathbf{r}_s$ of the GP equation Eq.(\ref{GP}) at which
$\rho(\mathbf{r}_s, t_\ast)=\infty$ for some fixed $t_\ast$ in
terms of $q$-trajectories, describes either the caustic  or focal
points of the trajectories. Notice that in the case of {\it linear
} Schr\"odinger equation the probability density $\rho$  is always
finite and caustics do not exist.

\begin{figure}
\includegraphics[width=70mm]{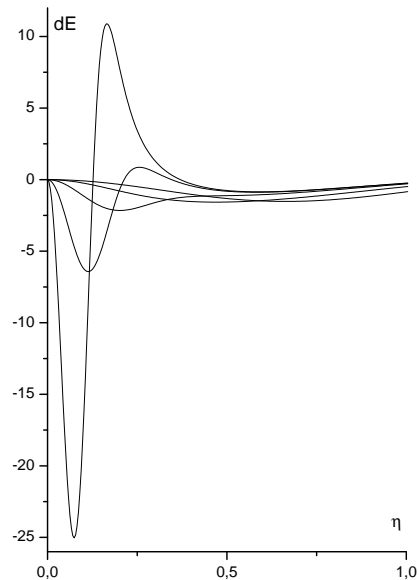}
\caption{\label{fig:Fig.1}  Energy$/N$ per sphere of radius
$|\bm{\eta}|$ at different times for the spherically symmetric
case. Greater peaks correspond to the moments of time closer to
$t_\ast$, $\bm{\eta}$ labels the initial position of the particle
in the flux of particles. Dimensionless units are defined in the
text.}
\end{figure}

The Newton equations Eqs.(\ref{EqNewton}) define the flux of
particles.  The appearance of the burst atoms in the mean field
theory can be qualitatively understood through analysis of the
energy distribution near the singularity in the  flux of particles
as shown in Figure~\ref{fig:Fig.1}. In this picture the peak with
positive energy density arises due to the acceleration of the
particles by the singularity.  This acceleration is proportional
to $g\nabla\rho$ and results in increase of the kinetic energy of
the particles surrounding the singularity. In
Figure~\ref{fig:Fig.1} the curves with greater peaks correspond to
the moments of time closer to $t_\ast$. The burst atoms can be
formed only if the dissipation losses occur faster than the
collapse domain is refilled by the flux. In this case one observes
intermittent implosions as reported in Ref.\cite{saito2001}.
The burst atoms cannot appear in the framework of the mean field
theory if the damping parameter is chosen to be large enough (cf.
Ref.\cite{abhi2002}). In this case the density cannot increase to
sufficiently large values and the peak of the positive energy
density is not formed.

The dynamics of the $q$-trajectories which begin at the points
located far from the origin in the $\bm{\eta}$-space cannot be
significantly affected by possible dissipative term. This is
because the compression time   of the breathing mode of the
harmonic oscillator, which is $1/4$ of the period of the
oscillator, is of the same order of magnitude as the time of the
collapse including the time of dissipation. This means that the
remote trajectories cannot reach the collapse domain.

In the presence of the burst atoms the condensate dynamics outside
the dissipation  domain is a quite complicated interference of the
'outgoing' burst atoms and 'incoming' atoms. Should we neglect
this interference, the fraction of the remaining atoms could be
found through the analysis of the particles flux falling into the
singularity and described by the Newton equations
Eqs.(\ref{EqNewton}). Notice that these equations are still valid
in the presence of  dissipative term. However in this case the
relation Eq.~(\ref{relation}) does not hold.

The analysis of flux of particles falling into the singularity
shows that the fraction of  atoms remaining in the condensate is
consistently    constant for the wide range of parameters. This
effect is supported by an experimental evidence~\cite{donley2001}.
In this work we deal with the general case of asymmetric trap. In
our previous work Ref.~\cite{rybin2001}  we reported for the
symmetric trap an effect similar to the described above. Our
computation was based on a Gaussian trial wave function which was
used to estimate a part of atoms focusing on the caustic. Here we
discuss different types of collapse which can exist in asymmetric
trap.

The paper is organized as follows. In the next section we discuss
different types of collapse which can occur in the asymmetric
trap. In the section 3 we obtain the collapsing fraction and show
that this fraction in the broad range of parameters consistently
remains almost constant.

\section{Breathing mode of the condensate}
In this  section we consider  types of collapse which can be
observed  in an  asymmetric trap. We use as the initial condition
 the Gaussian profile.
 For this profile
$$\rho_0=\frac{N}{\pi^{3/2}}\prod_{i=1}^3\frac{1}{{\rm
w}_i} e^{-x_i^2/{\rm w}_i^2}$$ the harmonic oscillator equation
has explicit solution, viz.
\begin{equation}
r_{i}({\bm\eta},t)=\sigma_{i}(t)\eta_i,\;\;\; \label{Eqho}
\end{equation}
where $$\sigma_{i}(t)=\sqrt{\cos^2(\omega_i t)+\frac{a_i^4}{{\rm
w}_i^4}\sin^2(\omega_i
t)},\;\;a_i=\sqrt{\frac{\hbar}{m\omega_i}}.$$ It seems natural, at
least for Gaussian initial conditions, to expect that the
breathing mode described by Eq.~(\ref{Eqho}) gives the main
contribution into the action
$$S({\mathbf r})=\int dt\int d{\bm\eta}\rho_0({\bm\eta}){\cal
L},$$ where $${\cal
L}=\sum_{i=1}^3\frac{m}{2}\dot{r}_i^2-\frac{\hbar^2}{2m\rho}(\nabla\sqrt{\rho})^2-
V-\frac{g}{2}\rho.$$

This provides us with the motivation to employ the variational
principle and to seek for the solution of the GP equation
Eq.~(\ref{GP}) in the form $r_{i}({\bm\eta},t)=\tau_{i}(t)\eta_i$.
Substituting this into the action
 we obtain
\begin{equation}
\ddot{\tau}_i+\omega_i^2\tau_i-\frac{\hbar^2}{m^2}\frac{1}{\tau_i^3}-
\frac{gN}{(2\pi)^{3/2}m}\frac{1}{\prod_k\tau_k}\frac{1}{\tau_i}=0
\label{Eqtau}
\end{equation}
The density as obtained through the relation Eq.~(\ref{relation})
reads
\begin{equation}
\rho(x,t)=\frac{N}{\pi^{3/2}}\prod_{i=1}^3\frac{1}{\tau_i}
e^{-x_i^2/\tau_i^2}\label{gauss} \end{equation} Here we have
included $\rm w_i$ in $\tau_i$ and  used $\rho_0$ with ${\rm
w_i}=1$.

The  result Eq.~(\ref{gauss}) is well-known  Gaussian trial wave
function Ref.\cite{gar}. This function describes  the breathing
(monopole) mode of the condensate.

In what follows we measure the time in units of $1/\omega$, where
$\omega=(\omega_x\omega_y\omega_z)^{1/3}$, and the distance in
units of the the oscillator length $a_{HO}=\sqrt{\hbar/m\omega}$.
For the axial symmetric equation we have
$\omega_x=\omega_y=\omega_r$ and $\tau_x=\tau_y=\tau_r$. Then the
system Eqs.(\ref{Eqtau}) is reduced to the following equations
\begin{equation}\label{rad1}
\ddot{\tau}_z+\beta^2\tau_z-\frac{1}{\tau_z^3}-\sqrt{\frac{2}{\pi}}\frac{\kappa}{\tau_z^2\tau_r^2}=0
\label{tauz}
\end{equation}
\begin{equation}
\ddot{\tau}_r+\frac{1}{\beta}\tau_r-\frac{1}{\tau_r^3}-\sqrt{\frac{2}{\pi}}\frac{\kappa}{\tau_z\tau_r^3}=0
\label{taur}
\end{equation}
Here $\beta=\frac{\omega_z}{\omega}$ and
$\kappa=\frac{Na_s}{a_{HO}}.$

For negative $\kappa$, such that $|\kappa|$ is greater than
certain critical value, the system of Eqs.(\ref{tauz}),
(\ref{taur}) has singular solutions with $\tau_r\to0$ as $t\to
t_\ast$. We found three different regimes of  collapse.

{\it Spherically} symmetric collapse is characterized by
$\tau_z\sim~\tau_r$ as $t\sim t_\ast$. The functions $\tau_r,
\tau_z$ near the singularity are described by the power law
$$\tau_z\sim\tau_{\ast}(t_\ast-t)^{2/5},\;\;\tau_r\sim\tau_{\ast}(t_\ast-t)^{2/5}$$
where $\tau_\ast$ is some constant. In this regime the compression
of the condensate is uniform in all directions.

{\it Radial} collapse is effectively two dimensional collapse for
which the condensate contracts only in the radial direction. The
behavior of $\tau_z, \tau_r$ near $t_\ast$  is defined by
\begin{equation}\label{rad1a}\tau_z\sim\lambda,\;\;\tau_r\sim\tau_{r\ast}(t_\ast-t)^{1/2}.\end{equation}
Here $\lambda$ is a constant and
$\tau_{r\ast}=\sqrt{2}\left(\sqrt{\frac{2}{\pi}}\frac{|\kappa|}{\lambda}-1\right)^{1/4}.$

 {\it Axial} collapse is characterized by $\tau_z\sim~\tau_r^2$ as
$t\sim t_\ast$. In this case the compression of the condensate is
faster in the axial direction. The function $\tau_z$ approaches
zero through an oscillating regime. This regime cannot be
described by a power law.

The actual type of  singularity depends on the initial conditions
for Eqs.(\ref{tauz}),(\ref{taur}). To begin the analysis we
specify the parameters and initial data as $\beta=0.5325,
\dot{\tau}_z(0)=0, \dot{\tau}_r(0)=0, \tau_z(0)=\tau_{z0}$ and
$\tau_r(0)=\tau_{r0}$. The constants $\tau_{z0}, \tau_{r0}$ are a
stationary solution of the system of Eqs.(\ref{tauz}),
(\ref{taur}) with $\kappa=\kappa_{in}>0$. This choice of
parameters models experiments on Feshbach resonances when the
scattering length is instantly changed from positive to negative
values in the initial stage of the dynamics.

For accurate implementation of numerical simulation  we use the
following change of variables
$$\tau_z=e^{\nu_z(\theta)},\;\tau_r=e^{\nu_r(\theta)},\;\theta=\ln\frac{t_c}{t_c-t},$$
where $t_c$ is a constant. The functions $\nu_r, \nu_z$ are
regular in the interval $[0,+\infty)$ if $t_c\leq t_\ast$ and
acquire a singularity at some $\theta_s<+\infty$ if $t_c>t_\ast$.

\begin{figure}
\includegraphics[width=80mm]{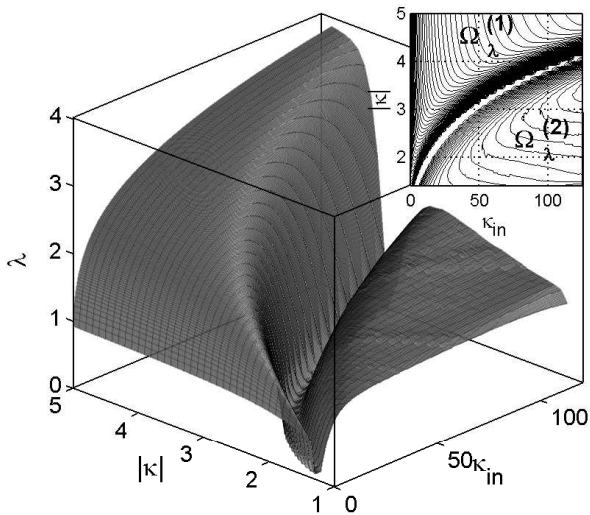}
\caption{\label{fig:Fig.3} The $3D$ plot and the contour plot of
$\lambda$ as a function of $|\kappa|$ and $\kappa_{in}$.}
\end{figure}

\begin{figure}
\includegraphics[width=80mm]{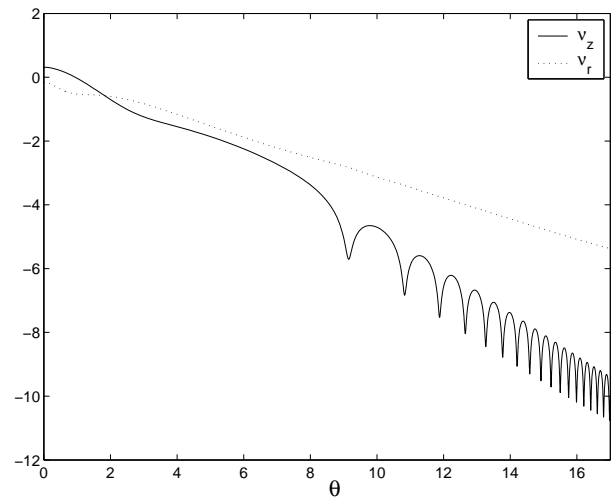}
\caption{\label{fig:Fig.2} Typical behavior of $\nu_z, \nu_r$ in
the axial collapse. Parameters are $|\kappa|=0.8$,
$\kappa_{in}=0$. This figure is derived in the limit $t_c\to
t_\ast-0$.}
\end{figure}

 We also
obtain $\lambda$ as a function of $\kappa, \kappa_{in}$ as shown
in Figure \ref{fig:Fig.3}. In this Figure the open domain
$\Omega_\lambda$ in $\kappa-\kappa_{in}$ plane consists of two
disconnected components $\Omega^{(1,2)}_\lambda$
$\left(\Omega_\lambda=\Omega^{(1)}_\lambda\bigcup
\Omega^{(2)}_\lambda\right)$ located on the different sides of the
groove. The overall domain $\Omega_\lambda$ corresponds to the
radial collapse. The component $\Omega^{(1)}_\lambda$ corresponds
to the simple radial collapse. In this subdomain the attraction
between atoms is strong, i.e. $|\kappa|$ is large.  This  results
in fast and monotonous decreasing of the Gaussian widths $\tau_i,
\,i=x,y,z$. Notice that while $\tau_{x,y}$ vanish, $\tau_{z}$
remains finite as in Eq.(\ref{rad1a}). In the second component of
$\Omega_\lambda$, i.e. in $\Omega^{(2)}_\lambda$ the density
$\rho(0,t)$ has at least one local maximum in the interval
$(0,t_\ast)$. In this regime the collapse occurs after a regular
compression-expansion cycle. The rates of contraction in the
radial and axial directions are alternating with  each other in
magnitudes. At some instances an intensive contraction in one
direction is compensated by {\it expansion} in the other. Notice
that the final stage of this process is {\it always} the
contraction in the {\it radial} direction. This is why we refer to
this type of collapse as to the {\it radial} collapse. We turn
next to the description of the open domain complimentary to the
domain $\Omega_\lambda$ in the $\kappa-\kappa_{in}$ plane. This is
the groove shown in Figure \ref{fig:Fig.3}. This domain
corresponds to the axial collapse. In the case of axial collapse
the ultimate stage of the dynamics is {\it always} the faster
contraction in the {\it axial} direction. A typical behavior in
axial collapse is shown in Figure \ref{fig:Fig.2}.

The boundary separating the open domain $\Omega_\lambda$ and the
groove, i.e. $\partial\Omega_\lambda$ corresponds to the
spherically symmetric collapse. It is interesting to notice that
even though the trap is asymmetric, the workings of nonlinear
dynamics still allow the spherical collapse to occur.

Intuitively, the existence of a regular cycle in dynamics of the
breathing mode  for $(\kappa, \kappa_{in})\in\Omega^{(2)}$
suggests that fewer  number of atoms must be removed to stabilize
the condensate. We further discuss this idea in the next section
where we analyze the flux of particles.

\section{The flux of particles}

The $q$-trajectories $r_{i}({\bm\eta},t)=\tau_i\eta_i$ derived in
the previous section focus at the symmetry axis when the time
approaches $t_\ast$ for any ${\bm\eta}$. However, for the large
$|{\bm\eta}|$ this is not true, since in this region the behavior
of the function ${\mathbf r}({\bm\eta},t)$  is determined by the
solution of the harmonic oscillator equation Eq.(\ref{Eqho}).

 The scope of analysis of equations for quantum trajectories
Eqs.(\ref{EqNewton}) proposed in this section attempts go beyond
the limitations posed by the Gaussian ansatz. In what follows we
construct an effective approximation for the flux of particles.
Our approximation effectively uses the notion of quantum
trajectories. We gain a motivation for this approach from the
analysis of the quasi-classical approximation for the
Schr\"odinger equation.  To construct the quasi-classical
approximation of the Schr\"odinger equation we have to solve the
Newton equations for the
 trajectories of particles, i.e. to solve $\ddot{r}_i=-\frac{\partial V}{\partial
r_i}$. The amplitude and the phase of the wave function are
determined through Eqs.(\ref{Eqforr}),(\ref{relation}).  We can
search further corrections substituting the quasi-classical
amplitude into the Newton equations Eqs.(\ref{EqNewton}) with
$g=0$. Iterating this procedure and assuming that iterations
converge we finally get the solution of the Schr\"odinger equation
in the time interval $[0,T]$ such that the function ${\mathbf
r}({\bm\eta},t)$ with $t\in[0,T]$ is invertible at each step. For
instance, in the case of harmonic oscillator and Gaussian initial
conditions this procedure converges in the interval defined by the
inequality $\prod_i\cos(\omega_it)>0$. In fact  we construct here
an iteration procedure $F$ of the form ${\mathbf
r}^{(n)}=F\left({\mathbf r}^{(n-1)}\right)$, where $n$ is the
number of iteration. These iterations start from the trajectories
of the classical harmonic oscillator
$r_{i}^{(0)}=r_{i}^{cl}({\bm\eta},t)=\cos(\omega_it)\eta_i$.
Notice that the mechanism of the iteration procedure $F$ is
disconnected from the quasi-classical approximation. This
procedure is
  equally well applicable
   in both linear and nonlinear ($g\ne 0$) cases. Indeed, solving Eqs.(\ref{EqNewton})
   in the nonlinear case, we can then find from
Eq.(\ref{relation}) a next iteration for the function $\rho$ which
we again  employ in Eqs.(\ref{EqNewton}), and so forth.

Below we  introduce an effective approximation for the quantum
trajectories as the first step of the iteration procedure $F$
described above. We assume that the starting  solution
Eq.(\ref{gauss}) delivered by the variational principle is
reasonably close to the exact solution.  Using
Eqs.(\ref{EqNewton}) with the density given by equation
Eq.(\ref{gauss}) and substituting Eq.(\ref{gauss}) into
Eqs.(\ref{EqNewton}) we obtain
\begin{equation}
\frac{\partial^2r_i}{\partial t^2}=-\frac{\partial
V_{eff}}{\partial r_i},\;r_i(0)=\eta_i,\;{\dot
r}_i(0)=0,\label{Eqeffective}
\end{equation}
where
$$V_{eff}(\mathbf{r},t)=\frac{1}{2}\sum_i\left(\frac{\omega_i^2}{\omega^2}
-\frac{1}{\tau_i^4}\right)r_i^2 +4\pi\kappa\rho(\mathbf{r},t)/N.$$
Notice that for the case of an axially symmetric trap it is
sufficient to consider the behavior of the $q$-trajectories in
$x$-$z$ plane. This can be achieved through reduction $r_y=0$.

At the moment of time $t_c$ close to $t_\ast$ we observe two main
types of trajectories: the trajectories which go away from the
center and the trajectories which tend to focus on the symmetry
axis. The domain in the ${\bm\eta}$-space corresponding to the
focussing  trajectories is denoted by $\Omega_s$. We now define
the collapsing fraction of atoms by an integral
\begin{equation}
N_e=\int\limits_{\Omega_s}\rho_0d\bm\eta.
\end{equation}
Here $\rho_0$ is the initial density.   It is physically plausible
to assume that the collapsing fraction is removed from the
condensate due to the dissipation. Then  the remaining fraction is
given by the formula $1-N_e/N$.

We analyze numerically the flux of particles falling into the
singularity. To do this we solve Eqs.(\ref{tauz}), (\ref{taur}),
(\ref{Eqeffective}).   At the moment $t=t_c$ the focussing
trajectories with reasonable accuracy get inside the domain
$\Omega_c$. This domain is restricted by the ellipsoid of
revolution of an ellipse with the semi $x$ and $z$-axes equal
$\tau_z(t_c)$ and $\tau_r(t_c)$ respectively. For $t_c\to t_\ast$
the domain $\Omega_c$ reduces to the interval $[-\lambda,\lambda]$
in the $z$-axis.

\begin{figure}
\includegraphics[width=85mm]{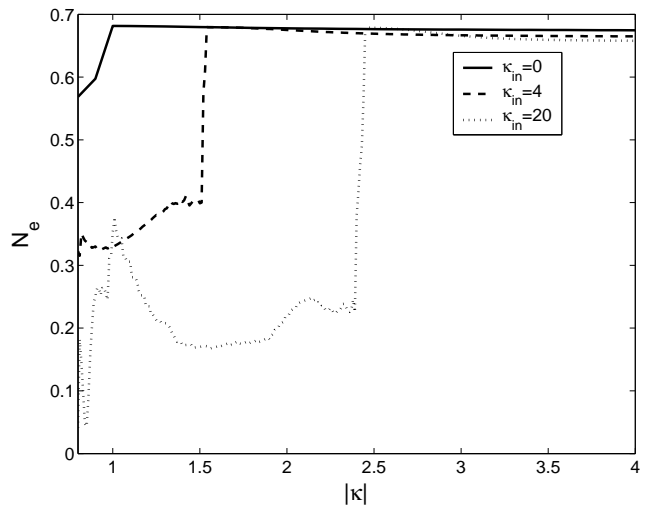}
\caption{\label{fig:Fig.4}The fraction $N_e$ as a function of
$|\kappa|$ for three different values of $\kappa_{in}$. }
\end{figure}

 We now technically define the domain
$\Omega_s$  in the $\bm\eta$-space such that if
$\mathbf{r}({\bm\eta},t_c)\in\Omega_c$  then
${\bm\eta}\in\Omega_s$. The results of simulations are presented
in Figure \ref{fig:Fig.4}. We find that the fraction $N_e/N$ is
consistently around 70\% for large values of $|\kappa|$ and
depends neither on $\kappa_{in}$ nor on $N$. More exactly this is
true for the subdomain $\Omega^{(1)}_\lambda$ corresponding to the
simple radial collapse described above. The collapsing fraction
decreases abruptly for $|\kappa|<|\kappa_{c}|$, where $\kappa_{c}$
is a certain threshold value specific to the given value of the
parameter $\kappa_{in}$. These thresholds for three different
values of $\kappa_{in}$ are shown in Figure~\ref{fig:Fig.4}. As
was explained above, below the threshold $|k_c|$, in the domain
$\Omega^{(2)}_\lambda$ and in the groove, we observe oscillatory
dynamical patterns. In the course of these oscillations we observe
losses in the collapsing fraction. For large $\kappa_{in}$ and in
this region the fraction $N_e/N$ is found to be around 20\% .

\section{Conclusions}
In this paper we suggest a simple model Eq.~(\ref{Eqeffective}) of
the BEC collapse based on the dynamical properties of the GP
equation. Our model is independent of the damping parameter and a
microscopic mechanism of the atom ejection. In the framework of
this model it is impossible to   correctly interpret the burst
atoms fraction. However, the  simplicity of the model allows us to
study the problem in wide range of the parameters.

Using the notion of the flux of particles we define the collapsing
fraction $N_e$ as the part of atoms focusing at the symmetry axis.
This focusing results in infinite increase of the density. Because
of this we assume that total collapsing fraction is removed from
the condensate. We estimate that $N_e/N=0.7$ for sufficiently
large $\kappa$ and the remaining fraction in this case is
insensitive to the choice of parameters.   Interestingly, our
results show that changing the initial aspect ratio by
$\kappa_{in}$
 it is possible to move from the domain $\Omega^{(1)}_\lambda$ to the domain
 $\Omega^{(2)}_\lambda$ for the same interaction strength $\kappa$.
 This means that the fraction of remaining atoms can be
increased for the larger initial aspect ratio controlled by
$\kappa_{in}$. In our opinion this effect deserves experimental
investigation.

\section*{Acknowledgements}

This work has been supported by the Academy of Finland (Project
No. 44875). I.P.V. is grateful to the Centre for International
Mobility (CIMO) for financial support. G.G.V. is grateful to the
University of Jyv\"askyl\"a, Finland for financial support.

\end{document}